\documentclass[pra,aps,twocolumn,showpacs]{revtex4}
\usepackage{epsf}
\usepackage{amsmath}
\usepackage{amssymb}

\newcommand{\bfm}[1]{{\boldsymbol #1}}

\begin{document}

\title{Violation of Bell inequality and entanglement
of decaying Werner states\footnote{to appear in {\em Phys. Lett. A}
(2004)}}

\author{Adam Miranowicz\footnote{{\em E-mail address}: miran@amu.edu.pl
(A. Miranowicz)}}
\date{\today}

\affiliation{Nonlinear Optics Division, Physics Institute, Adam
 Mickiewicz University, 61-614 Pozna\'n, Poland}

\begin{abstract}
Bell-inequality violation and entanglement, measured by Wootters'
concurrence and negativity, of two qubits initially in Werner or
Werner-like states coupled to thermal reservoirs are analyzed
within the master equation approach. It is shown how this simple
decoherence process leads to generation of states manifesting the
relativity of two-qubit entanglement measures.
\end{abstract}


\maketitle

\pagenumbering{arabic}

\section{Introduction}

Quantum nonlocality, responsible for violation of Bell-type
inequalities \cite{bell,clauser}, and entanglement (inseparability)
are the fundamental resources of modern quantum-information theory
and still the most surprising features of quantum mechanics (see,
e.g., \cite{tittel}). It is therefore desirable to investigate the
degrees of the Bell-inequality violation and of the entanglement of
a quantum state not only in relation to efficiency of
quantum-information processing, but also to understand better
subtle aspects of the physical nature.

It is well-known that pure states \cite{gisin} or a mixture of two
Bell states violate Bell inequalities whenever they are entangled.
So, one could naively think that the only mixed states that do not
violate the Bell inequalities are separable states. However, Werner
\cite{werner} demonstrated the existence of entangled states which
do not violate any Bell-type inequality.  The standard two-qubit
Werner state is defined by \cite{werner}
\begin{eqnarray}
{\rho }_{Y}^{(p)}(0) &=& p|Y\rangle \langle Y|
+\frac{1-p}{4}I\otimes {I} \label{N01}
\end{eqnarray}
being a mixture for $p\in \langle 0,1\rangle$ of the singlet state
$|Y\rangle =(|01\rangle - |10\rangle)/\sqrt{2}$ and the separable
maximally mixed state, given by $I\otimes I$, where $I$ is the
identity operator of a single qubit. Definition (\ref{N01}) is
often generalized to include mixtures of any maximally entangled
state (MES) instead of the singlet state only. So, e.g., one can
analyze the Werner-like state defined by \cite{munro,ghosh,wei1}
\begin{eqnarray}
{\rho }_{X}^{(p)}(0) &=&p|X\rangle \langle X|+
\frac{1-p}{4}I\otimes I, \label{N02}
\end{eqnarray}
as a convex combination of the Bell state $|X\rangle=(|00\rangle +
|11\rangle )/\sqrt{2}$ and $I\otimes I$. The original Werner state
(\ref{N01}), contrary to (\ref{N02}), is invariant if both qubits
are subjected to same unitary transformation, say $U\otimes U$.
Nevertheless, for a given $p$, both states (\ref{N01}) and
(\ref{N02}) exhibit the same entanglement properties, thus
(\ref{N02}) is also referred to as the Werner state
\cite{munro,ghosh,wei1}. In addition to the fact that the Werner
states can be entangled without violating any Bell inequality for
some values of parameter $p$, they can still be used for
quantum-information processing including teleportation
\cite{popescu,lee}. Moreover, the Werner states, given by
(\ref{N01}) and (\ref{N02}), can be considered maximally entangled
mixed states of two qubits \cite{ishizaka,munro} in the sense that
their degree of entanglement cannot be increased by any unitary
operations, and they have the maximum degree of entanglement for a
given linear entropy (and vice versa).

We will study the effects of a lossy environment, modelled by
thermal reservoirs, on the Bell-inequality violation and on the
entanglement of the initial Werner and Werner-like states in the
quest for new states including those having different orderings
induced by two entanglement measures: concurrence $C$ and
negativity $N$ (defined in Sect. IV). The relativity of the
entanglement measures was first observed by Eisert and Plenio
\cite{eisert}. They showed numerically, using Monte Carlo
simulation, that the condition
\begin{equation}
C(\rho_1) < C(\rho_2) \Leftrightarrow N(\rho_1) < N(\rho_2)
\label{N03}
\end{equation}
can be violated by some two-qubit mixed states, although it is
satisfied if $\rho_1$ and $\rho_2$ are the Werner states or pure
states. Virmani and Plenio \cite{virmani} demonstrated that all
good asymptotic entanglement measures are either identical or
impose different orderings of quantum states. The problem of
relativity of the entanglement measures was also studied in Refs.
\cite{zyczkowski99,verstraete,zyczkowski02,wei1,wei2}. Here, in
particular, we present analytical examples of different orderings
imposed by the concurrence and negativity for two-qubit states,
which violate the Bell inequality to the same degree.

The Letter is organized as follows. In Sect. II, we discuss a
dissipative model and give a general solution for two decaying
qubits initially in the Werner states. A comparative study of the
Bell-inequality violation and entanglement of various decaying
states are given is Sects. III and IV, respectively. A final
comparison and conclusions are given in Sect. V.

\section{Model for loss mechanism}

We analyze evolution of two initially correlated qubits subjected
to dissipation modelled by their coupling to thermal reservoirs
(phonon baths) as described by the following Hamiltonian
\begin{equation}
{H}=H_{S}+\hbar \sum_{k,n}\Omega _{n}^{(k)}({b}_{n}^{(k)})^{\dag
}{b} _{n}^{(k)}+\hbar \sum_{k,n}[g_{n}^{(k)}{a}_{k}^{\dag
}{b}_{n}^{(k)}+\mathrm{ h.c.}]  \label{N04}
\end{equation}
which is the sum of Hamiltonians for the system, $H_{S}$,
reservoirs and the coupling between them, respectively.  It is a
prototype model, where qubits can be implemented in various ways,
e.g., by single-cavity modes restricted in the Hilbert space
spanned by the two lowest Fock states (see, e.g.,
\cite{giovannetti}). In (\ref{N04}), ${a}_{k} $ is the annihilation
operator for the $k$th ($k=1,2$) qubit at the frequency $\omega
_{k}$; ${b}_{n}^{(k)}$ is the annihilation operator for the $n$th
oscillator in the $k$th reservoir at the frequency $\Omega
_{n}^{(k)}$, and $g_{n}^{(k)}$ are the coupling constants of the
reservoir oscillators. We assume no direct interaction between the
qubits, thus the system Hamiltonian is simply given by $H_{S}=\hbar
\sum_{k=1}^{2}\omega _{k}{a}_{k}^{\dag }{a}_{k}$. The standard
master equation for the model reads as
\begin{eqnarray}
\frac{\partial }{\partial t}{\rho } &=&\frac{1}{i\hbar
}[{H}_{S},{\rho } ]+\sum_{k=1}^{2}\frac{\gamma
_{k}}{2}\{\bar{n}_{k}(2{a}_{k}^{\dag }{\rho }{a}
_{k}-{a}_{k}{a}_{k}^{\dag }{\rho }-{\rho }{a}_{k}{a}_{k}^{\dag })  \notag \\
&&+(\bar{n}_{k}+1)(2{a}_{k}{\rho }{a}_{k}^{\dag }-{a}_{k}^{\dag
}{a}_{k}{ \rho }-{\rho }{a}_{k}^{\dag }{a}_{k})\}  \label{N05}
\end{eqnarray}
where ${\rho }$ is the reduced density operator for the qubits,
$\gamma _{k}$ is the damping constant and $\bar{n} _{k}$ is the
mean number of thermal photons of the $k$th reservoir. The exact
solution of (\ref{N05}) for arbitrary initial conditions and
arbitrary-dimensional systems is well known. By confining our
analysis to the initial qubit states coupled to the quiet
reservoirs ($\bar{n}_{1} =\bar{n}_{2}=0$), the solution in the
computational basis $\{|00\rangle,|01\rangle ,|10\rangle
,|11\rangle \}$ in the interaction picture can compactly be given
as
\begin{equation}
\rho (t)=\left[
\begin{array}{cccc}
{h} & f_{1}^{\ast } & f_{2}^{\ast } & \sqrt{g_{1}g_{2}}\rho _{03} \\
f_{1} & h_{1} & \sqrt{g_{1}g_{2}}\rho _{12} & \sqrt{g_{1}}g_{2}\rho _{13} \\
f_{2} & \sqrt{g_{1}g_{2}}\rho _{21} & h_{2} & g_{1}\sqrt{g_{2}}\rho _{23} \\
\sqrt{g_{1}g_{2}}\rho _{30} & \sqrt{g_{1}}g_{2}\rho _{31} &
g_{1}\sqrt{g_{2}} \rho _{32} & g_{1}g_{2}\rho _{33}
\end{array}
\right]   \label{N06}
\end{equation}
where the elements of the initial density matrix $\rho (0)$ are
denoted by $ \rho _{2i+k,2j+l}\equiv \langle i,k|\rho
(0)|j,l\rangle$; $g_{k}=\exp (-\gamma _{k}t)$ and
\begin{eqnarray}
f_{k} &=&\sqrt{g_{3-k}}[(1-g_{k})\rho _{3,3-k}+\rho _{k0}],  \notag \\
h_{k} &=&g_{3-k}[(1-g_{k})\rho _{33}+\rho _{kk})],  \notag \\
{h} &=&1-(h_{1}+h_{2}+g_{1}g_{2}\rho _{33}).  \label{N07}
\end{eqnarray}
We will apply solution (\ref{N06}) to analyze the effect of
dissipation on the Bell-inequality violation and entanglement of
the initial Werner states.

\section{Bell-inequality violation}

We will study violation of the Bell inequality due to Clauser,
Horne, Shimony and Holt (CHSH) \cite{clauser}. In a special case of
two qubits in an arbitrary mixed state $\rho$, one can apply an
effective criterion for violating the Bell inequality:
\begin{equation}
|\mathrm{Tr}\,(\rho \,{\mathcal B}_{\mathrm{CHSH}})|\leq 2
\label{N08}
\end{equation}
where ${\mathcal B}_{\mathrm{CHSH}}$ is the Bell operator given by
\begin{equation}
{\mathcal B}_{\mathrm{CHSH}}=\mathbf{a}\cdot \bfm{\sigma }\otimes
(\mathbf{ b}+\mathbf{b}^{\prime })\cdot \bfm{\sigma
}+\mathbf{a}^{\prime }\cdot \bfm{\sigma }\otimes
(\mathbf{b}-\mathbf{b}^{\prime })\cdot \bfm{\sigma } \label{N09}
\end{equation}
with its mean value maximized over unit vectors
$\mathbf{a},\,\mathbf{a}^{\prime },\,\mathbf{b},
\,\mathbf{b}^{\prime }$ in $\Re ^{3}$. Moreover, $\bfm{\sigma}$ is
the vector of the Pauli spin matrices $\sigma _{1},\sigma
_{2},\sigma _{3}$, and scalar product $\mathbf{a} \cdot \bfm{\sigma
}$ stands for $\sum\limits_{j=1}^{3}a_{j}\sigma _{j}$. By noting
that any $\rho$ can be represented in the Hilbert-Schmidt basis as
\begin{equation}
\rho =\tfrac{1}{4}\left( I\!\otimes I+\mathbf{r}\cdot \bfm{\sigma }
\otimes I+I\otimes \mathbf{s}\cdot \bfm{\sigma }+\!\!\!\sum
\limits_{n,m=1}^{3}t_{nm}\,\sigma _{n}\otimes \sigma _{m}\right)
\label{N10}
\end{equation}
where $\mathbf{r},\mathbf{s}$ are vectors in $\Re ^{3}$, Horodecki
et al. \cite{horodecki1,horodecki2} proved that the maximum
possible average value of the Bell operator in the state $\rho $ is
given by
\begin{equation}
\max_{{\mathcal B}_{\mathrm{CHSH}}}\,\mathrm{Tr}\,(\rho \,{\mathcal
B}_{ \mathrm{CHSH}})=2\,\sqrt{M(\rho )}  \label{N11}
\end{equation}
in terms of $M(\rho )=\max_{j<k}\;\{u_{j}+u_{k}\}$, where $u_{j}$
$(\,j=1,2,3)$ are the eigenvalues of the real symmetric matrix
$U_{\rho }=T_{\rho }^{T}\,T_{\rho }$; $T_{\rho }$ is the real
matrix formed by the coefficients $t_{nm}=\mathrm{Tr}\,(\rho
\,\sigma _{n}\otimes \sigma _{m})$, and $T_{\rho }^{T}$ is the
transposition of $T_{\rho }$. Thus, the necessary and sufficient
condition for violation of inequality (\ref{N08}) by the density
matrix (\ref{N10}) and Bell operator (\ref{N09}) for some choice of
$\mathbf{a},\mathbf{a}^{\prime },\mathbf{b},\mathbf{b}^{\prime }$
reads as $M(\rho )>1$ \cite{horodecki1,horodecki2}. To quantify the
degree of the Bell-inequality violation one can use $M(\rho)$,
$2\,\sqrt{M(\rho)}$ (see, e.g., \cite{ghosh,jakob}), or $\max
\,\{0,\,M(\rho )-1\,\}$ (see, e.g., \cite{jakobczyk}). But we
propose to use the following quantity
\begin{equation}
B(\rho )\equiv \sqrt{ \max \,\{0,\,M(\rho )-1\,\}}   \label{N12}
\end{equation}
which has a useful property that for any two-qubit pure state it is
equal to the entanglement measures such as concurrence and
negativity, defined in the next section. As $M(\rho )\leq 2$, it
holds $B(\rho )\in \langle 0,1\rangle $, where $B(\rho )=1$
corresponds to the maximal violation of inequality (\ref{N08}) and
$B(\rho )=0$ for states admitting the local hidden variable model.
The larger value of $B(\rho )>0$ the greater violation of the Bell
inequality. Thus, $B(\rho )$ can be used to quantify the degree of
the Bell-inequality violation (BIV), and for short it will be
referred to as the {\em BIV degree}.

The BIV degrees for the initial Werner states ${\rho
}_{X}^{(p)}(0)$ and ${\rho }_{Y}^{(p)}(0)$ are the same and equal
to
\begin{equation}
B_{X}^{(p)}(0)=B_{Y}^{(p)}(0)=\max \{0,2p^{2}-1\}^{1/2}
\label{N13}
\end{equation}
implying that the states violate the Bell inequality iff
$1/\sqrt{2}<p\le 1$. By changing the parameter $p$ into
$\phi=(1-3p)/2$, (\ref{N13}) goes into another well-known form
(see, e.g., \cite{horodecki1,horodecki2}). The BIV degree of the
maximally entangled states ($p=1$) is also maximal and equal to one
as shown in the left panels of Figs. 1, 2 and 4 at $t=0$.

By applying the solution (\ref{N06}) one observes that the initial
Werner state ${\rho }_{X}^{(p)}(0)$ decays as follows
\begin{equation}
{\rho }_{X}^{(p)}(t)=\tfrac{1}{4}\left[
\begin{array}{cccc}
h^{(+)} & 0 & 0 & 2p\sqrt{g_{1}g_{2}} \\
0 & h_{1}^{(+)} & 0 & 0 \\
0 & 0 & h_{2}^{(+)} & 0 \\
2p\sqrt{g_{1}g_{2}} & 0 & 0 & (1+p)g_{1}g_{2}
\end{array}
\right]  \label{N14}
\end{equation}
where ${h}^{(+)} =(2-g_{1})(2-g_{2})+pg_{1}g_{2}$ and $h_{k}^{(+)}
=g_{3-k}[2-(1+p)g_{k}]$. By applying the Horodecki criterion one
finds the eigenvalues of $U$ to be $ u_{1,2}=p^{2}g_{1}g_{2}$ and
$u_{1,2}\geq u_{3}=[(1-g_{1})(1-g_{2})+pg_{1}g_{2}]^{2}$, which
implies that the BIV degree evolves as
\begin{equation}
B_{X}^{(p)}(t)=\max \{0,2p^{2}g_{1}g_{2}-1\}^{1/2}.  \label{N15}
\end{equation}
Examples of decays of $B_{X}^{(p)}(t)$ for $p=1$ and $0.8$ are
presented graphically by assuming that the damping constants are
the same, for $\gamma_1=\gamma_2=0.1$ in Figs. 1(a,b) and 2(a,b),
or different, for $\gamma_1=0.1$ and $\gamma_2=0$, in Fig. 4. If
the qubits are initially in the Werner state ${\rho }_{Y}^{(p)}(0)$
then the evolution of the density matrix is described by
\begin{equation}
{\rho }_{Y}^{(p)}(t)=\frac{1}{4}\left[
\begin{array}{cccc}
h^{(-)} & 0 & 0 & 0 \\
0 & h_{1}^{(-)} & - 2p\sqrt{g_{1}g_{2}} & 0 \\
0 & - 2p\sqrt{g_{1}g_{2}} & h_{2}^{(-)} & 0 \\
0 & 0 & 0 & (1-p)g_{1}g_{2}
\end{array}
\right]  \label{N16}
\end{equation}
where $ h^{(-)} =(2-g_{1})(2-g_{2})-pg_{1}g_{2}$ and $h_{k}^{(-)}
=g_{3-k}[2-(1-p)g_{k}]$. Applying the same procedure as for the
state ${\rho }_{X}(t),$ one can find the following eigenvalues of
$U$: $u_{1,2}=p^{2}g_{1}g_{2}$ and $u_{1,2}\geq
u_{3}=[1-(g_{1}+g_{2})+(1-p)g_{1}g_{2}]^{2}\}$. Although the third
eigenvalue differs from that for ${\rho }_{X}(t),$ the BIV degrees
$B_{X}(t)$ and $B_{Y}(t)$ decay in the same manner:
\begin{equation}
B_{Y}^{(p)}(t)=\max \{0,2p^{2}g_{1}g_{2}-1\}^{1/2}=B_{X}^{(p)}(t)
\label{N17}
\end{equation}
as shown, e.g., by solid curves in Figs. 2(a,b) and 4. Obviously,
by changing the sign in the definitions of the Bell states
$|X\rangle$ and $|Y\rangle$, one finds the same decay of the BIV
degree. Thus, one could conjuncture that all MESs decay in the same
way. We will show that this is not true by analyzing the following
initial state
\begin{equation}
|Z\rangle =\frac{1}{2}(|00\rangle +|01\rangle +|10\rangle
-|11\rangle )\equiv \frac{1}{\sqrt{2}}(|0,+\rangle +|1,-\rangle )
\label{N18}
\end{equation}
where $|\pm \rangle =(|0\rangle \pm |1\rangle )/\sqrt{2}$. State
$|Z\rangle$ is another MES, which has the BIV degree (and also the
concurrence and negativity) equal to one, and can be obtained from
$|X\rangle$ by applying locally the Hadamard transformation to the
second qubit. One can also define another Werner-like state as a
mixture of the MES $|Z\rangle$ and the maximally mixed state given
by ${I}\otimes {I}$ as follows ($0\le p\le 1$):
\begin{equation}
{\rho }_{Z}^{(p)}(0)=p|Z\rangle \langle Z|+\frac{1-p}{4}{I}\otimes
{I} \label{N19}
\end{equation}
for which the BIV degree is given by (\ref{N13}) as for the
standard Werner state. For this and other reasons concerning the
entanglement properties (see Sect. IV) being the same as for the
state (\ref{N01}), we shall simply refer to (\ref{N19}) as the
Werner state despite the fact that (\ref{N19}) is not $U\otimes U$
invariant. The thermal reservoirs cause the decay of ${\rho
}_{Z}^{(p)}(0)$ as follows
\begin{equation}
{\rho }_{Z}^{(p)}\!(t)=\tfrac{1}{4}\!\!\left[
\begin{array}{cccc}
{h'} & pg_{1}\sqrt{g_{2}} & p\sqrt{g_{1}}g_{2} & -p\sqrt{g_{1}g_{2}} \\
pg_{1}\sqrt{g_{2}} & g_{2}(2-g_{1}) & p\sqrt{g_{1}g_{2}} &
-p\sqrt{g_{1}}
g_{2} \\
p\sqrt{g_{1}}g_{2} & p\sqrt{g_{1}g_{2}} & g_{1}(2-g_{2}) &
-pg_{1}\sqrt{
g_{2}} \\
-p\sqrt{g_{1}g_{2}} & -p\sqrt{g_{1}}g_{2} & -pg_{1}\sqrt{g_{2}} &
g_{1}g_{2}
\end{array}
\right]   \label{N20}
\end{equation}
where ${h'}=4+g_{1}g_{2}-2(g_{1}+g_{2}).$ The eigenvalues $ u_{k}$
are $\{p^{2}g_{1}g_{2},\frac{1}{2}(v\pm
\sqrt{v^{2}-4p^{4}g_{1}^{3}g_{2}^{3} })\}$ implying the following
decay of the BIV degree
\begin{equation}
B_{Z}^{(p)}(t)=\max \{0,p^{2}g_{1}g_{2}+\frac{1}{2}(v+\sqrt{
v^{2}-4p^{4}g_{1}^{3}g_{2}^{3}})-1\}^{1/2}  \label{N21}
\end{equation}
where $v=(1-g_{1})^{2}(1-g_{2})^{2}+p^{2}g_{1}g_{2}(g_{1}+g_{2})$.
For more explicit comparison of expressions for $B_{X,Y}^{(p)}(t)$
and $B_{Z}^{(p)}(t)$, we find their short-time approximations up to
linear terms in time for $p>1/\sqrt{2}$ as follows:
\begin{equation}
B_{X}^{(p)}(t)=B_{Y}^{(p)}(t)=\max \{0,Q-(\gamma _{1}+\gamma
_{2})\frac{p^{2}}{Q}t+{\mathcal O}(t^{2})\}, \label{N22}
\end{equation}
\begin{equation}
B_{Z}^{(p)}(t)=\max \{0,Q-[5(\gamma _{1}+\gamma _{2})-|\gamma
_{1}-\gamma _{2}|]\frac{p^{2}t}{4Q}+{\mathcal O}(t^{2})\}
\label{N23}
\end{equation}
where $Q=\sqrt{2p^{2}-1}$.  By attentively comparing (\ref{N17})
and (\ref{N21}), we find that it holds
\begin{eqnarray}
B_{X}^{(p)}(t) =  B_{Y}^{(p)}(t) \ge B_{Z}^{(p)}(t) \label{N24}
\end{eqnarray}
for any evolution times. In a special case of short times, this
result immediately follows from (\ref{N22}) and (\ref{N23}). So,
for both the qubits coupled to the reservoir(s) ($\gamma_{1},
\gamma_{2}> 0$), the evolution of $B_{Z}^{(p)}(t)>0$ differs from
that of $B_{X,Y}^{(p)}(t)$ as shown, e.g., by solid and broken
curves in Fig. 2 (a,b). However, by assuming that only one of the
qubits is coupled to the reservoir, say $\gamma_1\neq 0$ and
$\gamma_2=0$, the BIV degree of ${\rho }_{Z}^{(p)}(t)$ decreases at
the same rate as that of ${\rho }_{X}^{(p)}(t)$ and ${\rho
}_{Y}^{(p)}(t)$ and all the three the BIV degrees are given by
\begin{eqnarray}
B_{X}^{(p)}(\gamma _{2} &=&0,t)=B_{Y}^{(p)}(\gamma
_{2}=0,t)=B_{Z}^{(p)}(\gamma _{2}=0,t)  \notag \\
&=&\max \{0,2p^{2}g_{1}-1\}^{1/2} \label{N25}
\end{eqnarray}
which is a special case of (\ref{N17}) and (\ref{N21}). Clearly for
$\gamma _{2}=0$, (\ref{N23}) goes over into (\ref{N22}) as
expected. The decays of the BIV degrees for one of the damping
constants equal to zero are presented graphically by solid curves
in Fig. 4.

\begin{figure}
\epsfxsize=8cm\epsfbox{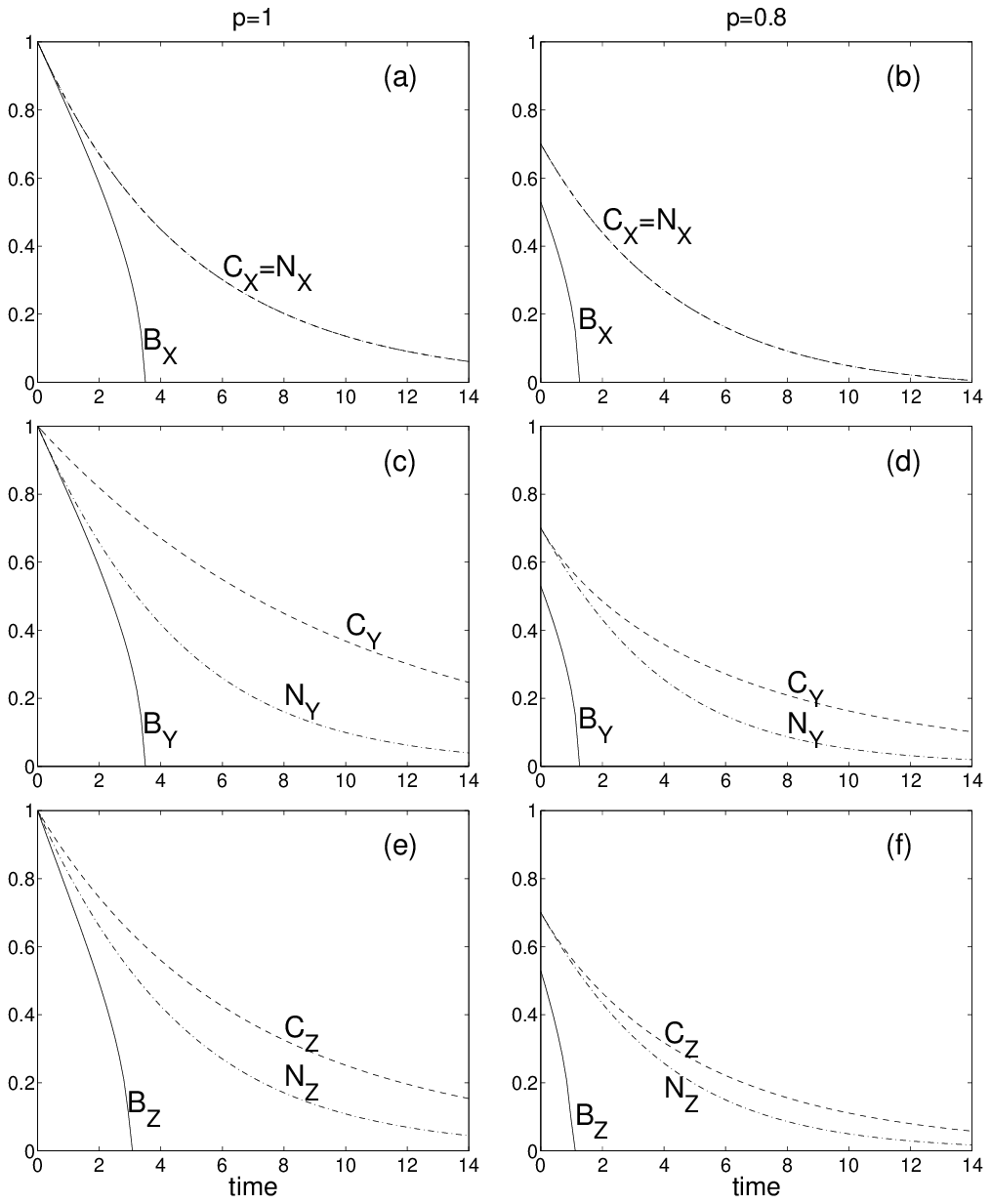}
 \caption{Decays of different measures for a given initial state in
each sub-figure: The Bell-inequality violation degree
$B_{\psi}^{(p)}(t)$, concurrence $C_{\psi}^{(p)}(t)$ and negativity
$N_{\psi}^{(p)}(t)$ for qubits initially in the Werner states
$\rho_{\psi}^{(p)}(0)$, where $\psi=X,Y,Z$ and $p=1$ (left panel)
or $p=0.8$ (right panel), coupled to the same reservoir with the
damping constant $\gamma_1=\gamma_2=0.1$.}
\end{figure}
\begin{figure}
\epsfxsize=8cm\epsfbox{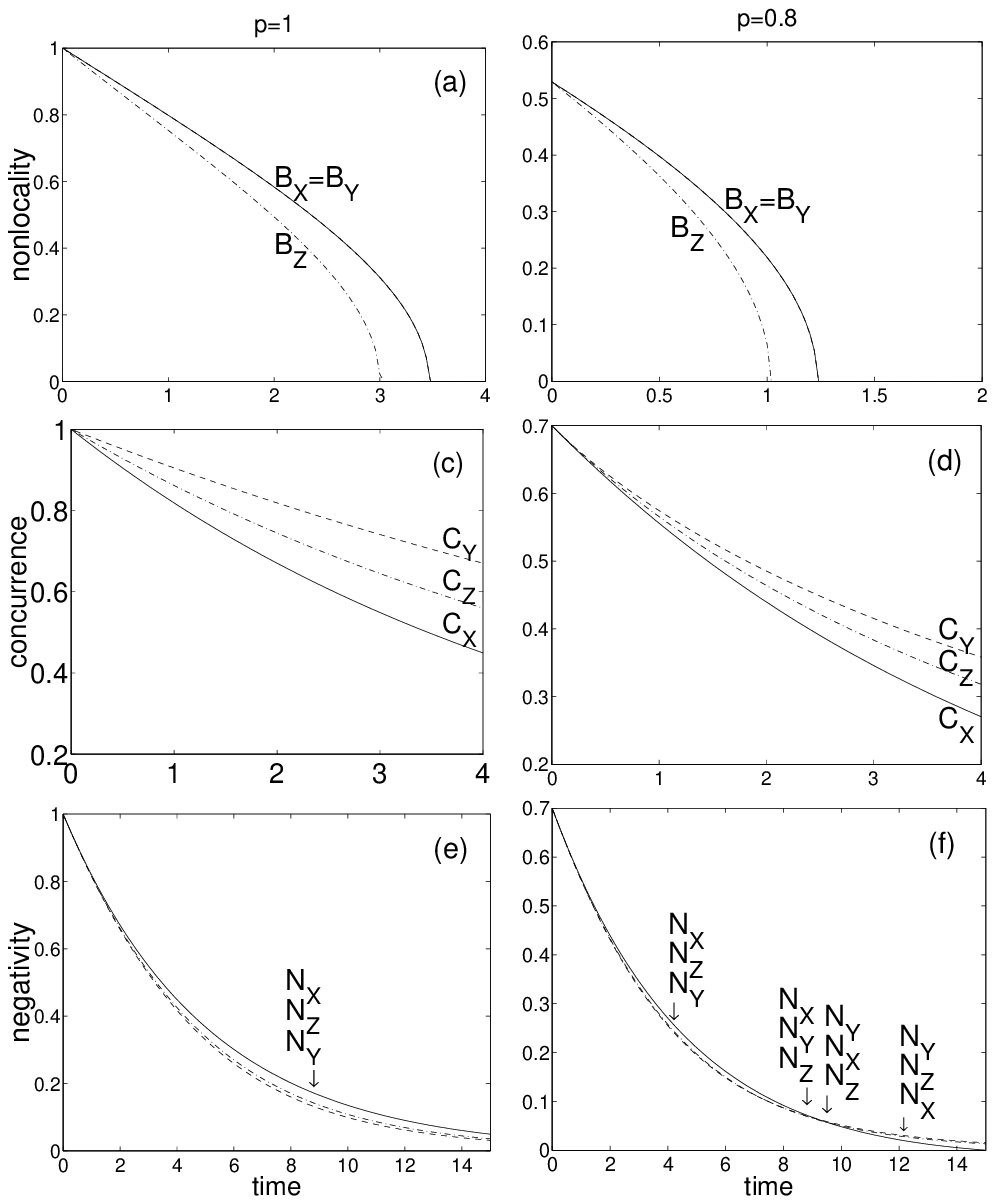}
 \caption{Decays of a given measure for different states in each
sub-figure. Notation and parameters are the same as in figure 1.}
\end{figure}
\begin{figure}
\hspace*{1mm} \epsfxsize=6cm\epsfbox{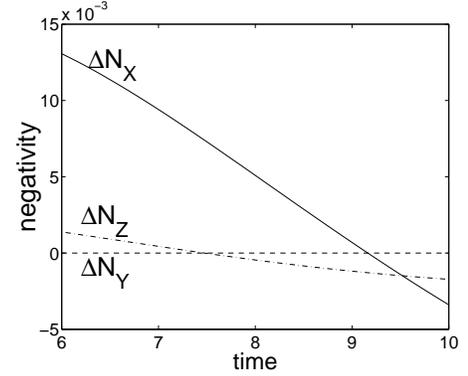}
\caption{Negativities $\Delta N_{\psi}^{(p)}(t)= N_{\psi}^{(p)}(t)-
N_{Y}^{(p)}(t)$ ($\psi=X,Y,Z$) corresponding to those presented in
Fig. 2(f) and Table I for the same $\gamma_1=\gamma_2=0.1$ and
$p=0.8$.}
\end{figure}
\begin{figure}
\epsfxsize=8cm\epsfbox{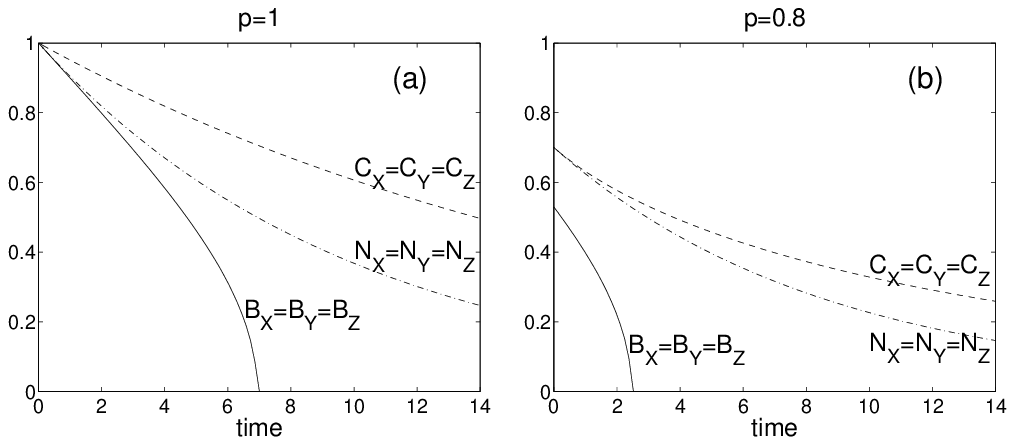}
 \caption{Same as in figure 1 but for one qubit coupled to the
reservoir with the damping constant $\gamma_1=0.1$, and the second
qubit undamped ($\gamma_2=0$).}
\end{figure}

\section{Entanglement}

To study the entanglement, we apply the concurrence and negativity
being related to the entanglement of formation and entanglement
cost, respectively.

The entanglement of formation of a mixed state $\rho $ is the
minimum mean entanglement of an ensemble of pure states $|\psi
_{i}\rangle $ that represents $\rho $ \cite{bennett2}:
\begin{equation}
E_{F}({\rho })=\min_{\{p_{i},|\psi _{i}\rangle
\}}\sum_{i}p_{i}E(|\psi _{i}\rangle \langle \psi _{i}|) \label{N26}
\end{equation}
where $\rho =\sum_{i}p_{i}|\psi _{i}\rangle \langle \psi _{i}|$ and
$E(|\psi _{i}\rangle \langle \psi _{i}|)$ is the entropy of
entanglement of pure state $|\psi _{i}\rangle $ defined by the von
Neumann entropy. As shown by Wootters \cite{wootters},  the
entanglement of formation for two qubits in an arbitrary mixed
state ${\rho }$ can explicitly be given as
\begin{equation}
E_{F}({\rho })=H\left(\textstyle{\frac{1}{2}}[1+\sqrt{1-C({\rho
})^{2}}]\right) \label{N27}
\end{equation}
where $H(x)$ is the binary entropy and $C({\rho })$ is the Wootters
concurrence defined by
\begin{equation}
C({\rho })=\max \{0,2\max_{i}\lambda _{i}-\sum_{i=1}^{4}\lambda
_{i}\} \label{N28}
\end{equation}
where $\lambda _{i}$ are the square roots of the eigenvalues of the
matrix ${\rho }({\sigma }_{y}\otimes {\sigma }_{y}){\rho }^{\ast
}({ \sigma }_{y}\otimes {\sigma }_{y})$, where ${\sigma }_{y}$ is
the Pauli spin matrix and the asterisk denotes complex conjugation.
$E_{F}(\rho)$ and $C(\rho)$ are monotonic functions of one another
and both range from 0 (for a separable state) to 1 (for a maximally
entangled state), so that ``one can take the concurrence as a
measure of entanglement in its own right'' \cite{wootters}.

The negativity is another measure of bipartite entanglement being
related to the Peres-Horodecki criterion \cite{peres,horodecki} and
defined by \cite{zyczkowski,eisert,vidal}
\begin{equation}
{N}({\rho })=\max \{0,-2\sum_{j}\mu _{j}\}  \label{N29}
\end{equation}
where ${\rho \equiv \rho }_{AB}$ is the density matrix of two
subsystems (say, $A$ and $B$ with $d_{A}$ and $d_{B}$ levels,
respectively), and the sum is taken over the negative eigenvalues
$\mu _{j}$ of the partial transpose ${\rho }^{T_{A}}$ of ${\rho }$
with respect to one of subsystems (say $A$) in the basis
$\{|0\rangle ,|1\rangle ,...,|d_{A}\rangle \}$:
\begin{equation}
{\rho }^{T_{A}}=\sum_{i,j=0}^{d_{A}-1}
\sum_{k,l=0}^{d_{B}-1}\langle i,k|\rho |j,l\rangle |j,k\rangle
\langle i,l|. \label{N30}
\end{equation}
For two-qubit ($d_{A}=d_{B}=2$) states, the sum in (\ref{N29}) can
be skipped as ${\rho }^{T_{A}}$ has at most one negative eigenvalue
\cite{sanpera}. The negativity, especially in low-dimensional
systems ($2\otimes 2$ and $ 2\otimes 3$),  is a useful measure of
entanglement satisfying the standard conditions
\cite{eisert1,vidal}. The negativity (\ref{N29}) ranges from 0 (for
a separable state) to 1 (for a MES) similarly to the concurrence
and the BIV degree.  It is worth noting that the logarithmic
negativity, $\log _{2}[{N} ({\rho })+1]$, has a simple operational
interpretation as a measure of the entanglement cost for the exact
preparation of a two-qubit state $\rho$ under quantum operations
preserving the positivity of the partial transpose (PPT)
\cite{audenaert,ishizaka04}.

The concurrences and negativities for all the three initial Werner
states ${\rho }_{\psi}^{(p)}(0)$ ($\psi=X,Y,Z$) are the same and
equal to
\begin{equation}
C_{\psi}^{(p)}(0)={N}_{\psi}^{(p)}(0)=\max \{0,\tfrac{1}{2}(3p-1)\}
\label{N31}
\end{equation}
but different from their BIV degree $B_{\psi}^{(p)}(0)$, given by
(\ref{N13}). For the Bell states, all the entanglement measures and
the BIV degree are equal to one. However, by decreasing parameter
$p$, the BIV degree of the Werner states decreases faster than
their entanglement. For example of $p=0.8$, the initial values of
the concurrences and negativities are equal to 0.7 while the BIV
degree is 0.529 as shown in the right panels of Figs. 1, 2, and 4.
By comparing (\ref{N13}) and (\ref{N31}) it is seen that the Werner
states are entangled iff $1/3<p \le 1$. Thus, the Werner states for
$p\in (1/3,1/\sqrt{2}\rangle $ are entangled although admitting a
local hidden model, i.e., satisfying the Bell inequality
\cite{werner}.

Qubits initially in the Werner state ${\rho }_{X}^{(p)}(0)$ coupled
to the thermal reservoirs exhibit dissipation described by
(\ref{N14}). We find with the help of the Wootters formula that the
concurrence for ${\rho }_{X}^{(p)}(t)$ exhibits the following decay
\begin{equation}
C_{X}^{(p)}(t)=\max \{0,\frac{\sqrt{g_{1}g_{2}}}{2}\left(2p-
\sqrt{(2-qg_{1})(2-qg_{2})}\right) \}  \label{N32}
\end{equation}
with $q=1+p$. While the negativity, according to Peres-Horodecki
criterion applied for ${\rho }_{X}^{(p)}(t)$, decays as
\begin{eqnarray}
N_{X}^{(p)}(t) &=&\max \{0,\frac{1}{2}[-g_{1}-g_{2}+(1+p)g_{1}g_{2}
\notag \\
&&\qquad\qquad +\sqrt{(g_{1}-g_{2})^{2}+4p^{2}g_{1}g_{2}}]\}.
\label{N33}
\end{eqnarray}
By assuming that both qubits are coupled to the same reservoir
described the damping constant $\gamma \equiv \gamma _{1}=\gamma
_{2}$, we observe that the concurrence and negativity are the same
for all evolution times as described by
\begin{equation}
C_{X}^{(p)}(t)={N}_{X}^{(p)}(t)=\max
\{0,\frac{g}{2}[(1+p)g-2(1-p)]\}, \label{N34}
\end{equation}
where $g=\exp (-\gamma t)$, as clearly depicted in Fig. 1(a,b). In
another special case, for the initial Bell states ($p=1)$, the
solutions for the concurrence and negativity simplify to
\begin{eqnarray}
C_{X}^{(1)}(t) &=&\sqrt{g_{1}g_{2}}\left(
1-\sqrt{(1-g_{1})(1-g_{2})}\right)
,  \notag \\
{N}_{X}^{(1)}(t) &=&g_{1}g_{2},  \label{N35}
\end{eqnarray}
respectively.

On the other hand, for qubits decaying from the initial Werner
state ${\rho }_{Y}^{(p)}(0)$, as described by  (\ref{N16}), the
concurrence decays in time as
\begin{eqnarray}
C_{Y}^{(p)}(t) &=& \max \{0,\frac{1}{2}\sqrt{g_{1}g_{2}}( 2p -
\sqrt{1-p} \notag \\ && \qquad \times
\sqrt{(2-g_{1})(2-g_{2})-pg_{1}g_{2}})\} \label{N36}
\end{eqnarray}
while the negativity vanishes as follows
\begin{eqnarray}
N_{Y}^{(p)}(t) &=&\max \{0,\frac{1}{2}(-2+g_{1}+g_{2}-(1-p)g_{1}g_{2}  \notag \\
&&+\sqrt{(2-g_{1}-g_{2})^{2}+4p^{2}g_{1}g_{2}})\}.  \label{N37}
\end{eqnarray}
Note that decays of all the entanglement measures, similarly to the
BIV degree, are independent of the sign in definitions of
$|X\rangle$ and $|Y\rangle$. In a special case for qubits coupled
to the same reservoir ($ \gamma _{1}=\gamma _{2}$), (\ref{N36}) and
(\ref{N37}) simplify, respectively, to
\begin{eqnarray}
C_{Y}^{(p)}(t) &=&\max \{0,g \left(p -\sqrt{r(1-g)+\tfrac{1}{4}r^{2}g^{2}}\right)\}, \\
N_{Y}^{(p)}(t) &=&\max \{0,\sqrt{(1-g)^{2}+p^{2}g^{2}}-\tfrac{1}{2}
rg^{2}-(1-g)\}  \notag  \label{N38}
\end{eqnarray}
with $r=1-p$. In another special case for the initial Bell state
($p=1$), the entanglement measures (\ref{N36}) and (\ref{N37})
reduce to
\begin{eqnarray}
C_{Y}^{(1)}(t) &=&\sqrt{g_{1}g_{2}}, \label{N39} \\
N_{Y}^{(1)}(t) &=&\frac{1}{2}\left(
\sqrt{(2-g_{1}-g_{2})^{2}+4g_{1}g_{2}} +g_{1}+g_{2}-2\right),
\notag
\end{eqnarray}
respectively. It is worth mentioning that, in contrast to
$C_{X}^{(p)}(t) ={N}_{X}^{(p)}(t)$ for $\gamma _{1}=\gamma _{2},$
the evolutions $C_{Y}^{(p)}(t)$ and ${N}_{Y}^{(p)}(t)$ are
different. Finally, let us briefly discuss the effect of
dissipation of the initial Werner state ${\rho }_{Z}^{(p)}(0)$,
described by (\ref{N20}), on the entangled measures. The general
analytical formulas for the $p$-dependent concurrence and
negativity are quite lengthy thus are not presented here explicitly
although were used for plotting the corresponding curves in Figs.
1(e,f), 2(e,f) and 3. However, in a special case for the initial
Bell-like state ($p=1$) the decays of the entanglement measures are
simply given by:
\begin{eqnarray}
 \label{N40}
C_{Z}^{(1)}(t) &=&\sqrt{g_{1}g_{2}}\left( 1-\tfrac{1}{2}\sqrt{
(1-g_{1})(1-g_{2})}\right),   \\
{N}_{Z}^{(1)}(t)
&=&\tfrac{1}{2}\sqrt{g_{1}g_{2}(5+g_{1}g_{2}-2G)+4(1-G)^{2}} +G-1
\notag
\end{eqnarray}
with $G=(g_{1}+g_{2})/2$. By assuming the same damping constants
$\gamma _{1}=\gamma _{2}$, e.g., the concurrence formula reduces to
$C_{Z}^{(1)}(t) =g(1+g)/2$. A graphical comparison of the decays of
all the measures is shown in Fig. 2 and of the negativities in Fig.
3 and  Table I for $\gamma _{1}=\gamma _{2}=0.1$. Since the
differences between the negativities $N_{\psi}^{(p)}(t)$ are not
clear enough in Fig. 2(f), the curves were redrawn in Fig. 3 for
the rescaled negativities $\Delta N_{\psi}^{(p)}(t)\equiv
N_{\psi}^{(p)}(t)- N_{Y}^{(p)}(t)$ with $\psi=X,Y,Z$.

For clear analysis of the entanglement measures we will now focus
on the special case for $p=1$. A comparison of Eqs. (\ref{N35}),
(\ref{N39}), and (\ref{N40}) implies that the following
inequalities are satisfied
\begin{eqnarray}
C_{Y}^{(1)}(t) \ge  C_{Z}^{(1)}(t) \ge C_{X}^{(1)}(t) \label{N41}
\end{eqnarray}
for any evolution times. In the short time approximation, we find
that the concurrences for $\rho^{(1)}_{\psi}$ ($\psi=X,Y,Z$)  decay
up to linear terms in time as follows
\begin{eqnarray}
C_{\psi}^{(1)}(t)&=&1-\frac{1}{2}(\gamma _{1}+ \gamma _{2}+
f_{\psi}\sqrt{\gamma _{1}\gamma _{2}})t+{\mathcal O}(t^{2})
\label{N42}
\end{eqnarray}
where $f_X=2$, $f_Y=0$, and $f_Z=1$, which confirms the validity of
(\ref{N41}). Contrary to the concurrences, the negativities in the
short-time limit decay as follows
\begin{equation}
N_{\psi}^{(1)}(t)=1-(\gamma _{1}+ \gamma _{2})t+\frac{1}{2}(\gamma
_{1}^2 +f'_{\psi} \gamma _{1}\gamma _{2} +\gamma _{2}^2
)t^2+{\mathcal O}(t^{3}) \label{N43}
\end{equation}
where $f'_X=2$, $f'_Y=1$, and $f'_Z=5/4$, which implies that at
short evolution times the inequalities hold
\begin{eqnarray}
N_{X}^{(1)}(t) \ge  N_{Z}^{(1)}(t) \ge N_{Y}^{(1)}(t). \label{N44}
\end{eqnarray}
A closer look at the inequalities (\ref{N24}), (\ref{N41}) and
(\ref{N44}) enables us to conclude that there exist states of two
qubits (e.g., $\rho_{X}^{(1)}(t)$ and $\rho_{Y}^{(1)}(t)$ at some
short evolution time t) exhibiting the same BIV degree,
$B_{X}^{(1)}(t)=B_{Y}^{(1)}(t)$, but different entanglement
measures in such a way that the concurrence $C_{X}^{(1)}(t)$ is
smaller than $C_{Y}^{(1)}(t)$, while the negativity
$N_{X}^{(1)}(t)$ is greater than $N_{Y}^{(1)}(t)$. Obviously, the
inequalities (\ref{N41}) for the concurrences correspond to those
for the entanglement of formation, while the inequalities
(\ref{N44}) for the negativities correspond to those for the
PPT-entanglement cost. We stress that for longer times inequalities
different from (\ref{N44}) are satisfied for $p<1$ as presented in
Table I. By analyzing this Table, other states differently ordered
by the entanglement measures are readily recognized, including
those at time $t=t_2$, for which $B_{X}^{(.8)}(t)=B_{Y}^{(.8)}(t)$,
$N_{X}^{(.8)}(t) =N_{Y}^{(.8)}(t)$ but
$C_{X}^{(.8)}(t)<C_{Y}^{(.8)}(t)$.

By comparing the series expansions (\ref{N42}) and (\ref{N43}) for
$\gamma_1,\gamma_2>0$, we can conclude that all the three
negativities evolve in a more similar way (precisely they are the
same up to linear terms in time) in comparison to more distinct
evolutions of the corresponding concurrences. Also by analyzing
(\ref{N22}) and (\ref{N43}) for $p=1$, one observes that the
negativities $N_{X,Y,Z}^{(1)}(t)$ and the BIV degrees
$B_{X,Y}^{(1)}(t)$ decay in the same manner up to linear terms in
time for arbitrary values of $\gamma_1$ and $\gamma_2$. Other
similarities of the decays of the entanglement and/or the BIV
degree can be found for some special choices of the damping
constants. In particular, (\ref{N34}) is valid for
$\gamma_1=\gamma_2$. In another special case of only one of the
qubits coupled to the reservoir, e.g. $\gamma_1\neq 0$ and
$\gamma_2=0$, the BIV degrees of all the three states decrease, as
given by (\ref{N25}), but also the entanglement measures decrease
at the same rate as expressed by their concurrence
\begin{eqnarray}
C_{X}^{(p)}(\gamma _{2} =0,t)=C_{Y}^{(p)}(\gamma
_{2}=0,t)=C_{Z}^{(p)}(\gamma _{2}=0,t)  \notag \\
=\max \{0,\frac{1}{2}\sqrt{g_{1}}\left(
2p-\sqrt{(1-p)[2-(1+p)g_{1}]} \right) \}  \label{N45}
\end{eqnarray}
and negativity
\begin{eqnarray}
N_{X}^{(p)}(\gamma _{2} =0,t)=N_{Y}^{(p)}(\gamma
_{2}=0,t)=N_{Z}^{(p)}(\gamma _{2}=0,t)  \notag \\
=\max \{0,\frac{1}{2}\left( pg_{1}+\sqrt{(1-g_{1})^{2}+4p^{2}g_{1}}
-1\right) \}.  \label{N46}
\end{eqnarray}
By restricting to the case of the initial MESs ($p=1$), the above
equations reduce to ($\psi =X,Y,Z$):
\begin{eqnarray}
C_{\psi }^{(1)}(\gamma _{2}=0,t)=\sqrt{g_{1}}, \notag \\
N_{\psi }^{(1)}(\gamma _{2}=0,t)=g_{1}.
 \label{N47}
\end{eqnarray}
These conclusions are confirmed numerically on the examples of
$\gamma_1=0.1$, $p=0.8$ and $p=1$ as shown in Fig. 4.

\begin{table}
\caption{Comparison of the negativities for the three initial
Werner states for $p=0.8$, $\gamma_1=\gamma_2=0.1$ as in Figs. 2(f)
and 3. The characteristic evolution times are $t_1\approx7.4745$,
$t_2\approx 9.1613$, $t_3\approx9.5209$. The corresponding
inequalities for the BIV degrees are $B_X^{(p)}(t)=B_Y^{(p)}(t)\ge
B_Z^{(p)}(t)$, and for the concurrences are $C_Y^{(p)}(t)\ge
C_Z^{(p)}(t)\ge C_X^{(p)}(t)$.}
\begin{center}
\begin{tabular}{l l}
time & negativities
\\ \hline
$t\in(0,t_1)$       & $N_X^{(p)}(t)>N_Z^{(p)}(t)>N_Y^{(p)}(t)$ \\
$t=t_1$         & $N_X^{(p)}(t)>N_Y^{(p)}(t)=N_Z^{(p)}(t)$ \\
$t\in(t_1,t_2)$     & $N_X^{(p)}(t)>N_Y^{(p)}(t)>N_Z^{(p)}(t)$ \\
$t=t_2$         & $N_X^{(p)}(t)=N_Y^{(p)}(t)>N_Z^{(p)}(t)$ \\
$t\in(t_2,t_3)$     & $N_Y^{(p)}(t)>N_X^{(p)}(t)>N_Z^{(p)}(t)$ \\
$t=t_3$         & $N_Y^{(p)}(t)>N_X^{(p)}(t)=N_Z^{(p)}(t)$ \\
$t\in(t_3,100)$  & $N_Y^{(p)}(t)>N_Z^{(p)}(t)>N_X^{(p)}(t)$ \\
\\
\end{tabular}
\end{center}
\end{table}

\section{Conclusions}

We have analyzed quantum-information properties of two decaying
optical qubits prepared initially in Werner or Werner-like states
and coupled to thermal reservoirs within the master equation
approach. We have studied in detail a degree of violation of the
Bell inequality due to Clauser, Horne, Shimony and Holt
\cite{clauser} by applying a parameter $B$ related to the maximum
possible mean value of the Bell operator in a given state according
to the Horodecki criterion \cite{horodecki1}. On the other hand,
the degree of the entanglement was expressed by the Wootters
concurrence $C$ \cite{wootters}, as a measure of the entanglement
of formation, and by the negativity $N$ based on the
Peres-Horodecki criterion \cite{peres,horodecki} and related to the
PPT-entanglement cost \cite{audenaert}. We have observed, as
manifestations of the symmetry of our particular decoherence
mechanism, the following properties of the decaying Werner states
in relation to the Bell-inequality violation degree $B$ and the
entanglement measures $C$ and $N$: If only one qubit is coupled to
the thermal reservoir than those decays are independent of the
initial Werner or Werner-like state for a given $p$. However, if
both qubits are coupled to the reservoir(s) then the decays of $C$
and $N$, and in some cases $B$, depend on the initial Werner or
Werner-like state. By analyzing these decays, we have found states
(say $\rho$ and $\sigma$) of two qubits exhibiting the same degree
the Bell-inequality violation, $B_{\rho}=B_{\sigma}$, but different
entanglement measures in such a way that the concurrence $C_{\rho}$
is smaller than $C_{\sigma}$, while the negativity $N_{\rho}$ is
greater than $N_{\sigma}$. We have also found other states $\rho$
and $\sigma$, for which either (i) $B_{\rho}=B_{\sigma}$,
$C_{\rho}<C_{\sigma}$ and $N_{\rho}<N_{\sigma}$, or (ii)
$B_{\rho}=B_{\sigma}$, $N_{\rho}= N_{\sigma}$ and
$C_{\rho}<C_{\sigma}$.  Thus, the analysis of the decaying Werner
states shows clearly the relativity of two-qubit entanglement
measures.

\section*{Acknowledgments}

The author thanks Jens Eisert, Andrzej Grudka, Pawe\l{}~ Horodecki,
Wies\l{}aw Leo\'nski, and Ryszard Tana\'{s} for stimulating
discussions.


\begin{thebibliography}{99}

\bibitem{bell}
J. S. Bell, {Physics} (Lon Island City, N.Y.) {\bf 1}, 195 (1964).

\bibitem{clauser} J. F. Clauser, M. A. Horne, A. Shimony, R. A. Holt,
Phys. Rev. Lett. {\bf 23}, 880 (1969).

\bibitem{tittel}
W. Tittel,  G. Weihs, Quantum Inform. Comp. {\bf 1}, 3 (2001).

\bibitem{gisin}
N. Gisin, Phys. Lett. A {\bf 154}, 201 (1991).

\bibitem{werner}
R. F. Werner, {Phys. Rev. A} {\bf 40}, 4277 (1989).

\bibitem{munro}
W. J. Munro, D. F. V. James, A. G. White, P. G. Kwiat, {Phys. Rev.
A} {\bf 64}, 030302(R) (2001).

\bibitem{ghosh}
S. Ghosh, G. Kar, A. Sen De,  U. Sen, Phys. Rev. A {\bf 64}, 044301
(2001).

\bibitem{wei1}
T. C. Wei, K. Nemoto, P. M. Goldbart, P. G. Kwiat, W. J. Munro, F.
Verstraete, Phys. Rev. A {\bf 67}, 022110 (2003).

\bibitem{popescu}
S. Popescu, Phys. Rev. Lett. {\bf 72}, 797 (1994).

\bibitem{lee}
J. Lee, M. S. Kim, Phys. Rev. Lett. {\bf 84}, 4236 (2000).

\bibitem{ishizaka}
S. Ishizaka,  T. Hiroshima, {Phys. Rev. A} {\bf 62}, 022310 (2000).

\bibitem{eisert}
J. Eisert,  M. Plenio, J. Mod. Opt. {\bf 46}, 145 (1999).

\bibitem{virmani}
S. Virmani,   M. B. Plenio, Phys. Lett. A {\bf 268}, 31 (2000).

\bibitem{zyczkowski99}
K. \.Zyczkowski, Phys. Rev. A {\bf 60}, 3496 (1999).

\bibitem{verstraete}
F. Verstraete, K. M. R. Audenaert, J. Dehaene,  B. De Moor, J.
Phys. A {\bf 34}, 10327 (2001).

\bibitem{zyczkowski02}
K. \.Zyczkowski,  I. Bengtsson, Ann. Phys. (N.Y.) {\bf 295}, 115
(2002).

\bibitem{wei2}
T. C. Wei,  P. M. Goldbart, Phys. Rev. A {\bf 68}, 042307 (2003).

\bibitem{giovannetti}
V. Giovannetti, D. Vitali, P. Tombesi,  A. K. Ekert, {Phys. Rev. A}
{\bf 62}, 032306 (2000).

\bibitem{horodecki1} R. Horodecki, P. Horodecki,  M. Horodecki, Phys.
Lett. A {\bf 200}, 340 (1995)

\bibitem{horodecki2}
R. Horodecki, Phys. Lett. A {\bf 210}, 223 (1996).

\bibitem{jakob}
M. Jakob, Y. Abranyos,  J. A. Bergou, {Phys. Rev. A} {\bf 66},
022113 (2002).

\bibitem{jakobczyk} L. Jak\'obczyk,  A. Jamr\'oz, Phys. Lett.
A {\bf 318}, 318 (203).

\bibitem{bennett2}
C. H. Bennett, D. P. DiVincenzo, J. A. Smolin,  W. K. Wootters,
{Phys. Rev. A} {\bf 54}, 3824 (1996).

\bibitem{wootters}
W. K. Wootters, {Phys. Rev. Lett.} {\bf 80}, 2245 (1998).

\bibitem{peres}
A. Peres, {Phys. Rev. Lett.} {\bf 77}, 1413  (1996).

\bibitem{horodecki}
M. Horodecki, P. Horodecki,  R. Horodecki, {Phys. Lett. A} {\bf
223}, 1  (1996).

\bibitem{zyczkowski}
K. \.Zyczkowski, P. Horodecki, A. Sanpera,  M. Lewenstein, {Phys.
Rev. A} {\bf 58} 883 (1998).

\bibitem{eisert1}
J. Eisert, {Ph.D. Thesis} (University of Potsdam) (2001).

\bibitem{vidal}
G. Vidal,  R. F Werner, {Phys. Rev. A} {\bf 65}, 032314 (2002).

\bibitem{sanpera}
A. Sanpera, R. Tarrach,  G. Vidal, {Phys. Rev. A} {\bf 58}, 826
(1998).

\bibitem{audenaert}
K. Audenaert, M. B. Plenio,  J. Eisert, {Phys. Rev. Lett.} {\bf
90}, 27901 (2003).

\bibitem{ishizaka04}
S. Ishizaka, Phys. Rev. A {\bf 69}, 020301 (2004).

\end{thebibliography}
\end{document}